\documentclass[12pt,epsfig,citesort]{article}
\usepackage{geometry}
\newcommand{\ice}[1]{\relax}
\usepackage{graphicx}
\usepackage{epsfig}
\usepackage{latexsym}
\usepackage{amsmath}
\usepackage{amsfonts}
\usepackage{amsxtra}

\def\be{\begin{equation}}
\def\ee{\end{equation}}
\def\bea{\begin{eqnarray}}
\def\eea{\end{eqnarray}}

\def\ap#1#2#3   {{ Ann. Phys. (NY)} {\bf#1} (#2) #3.}
\def\apj#1#2#3  {{  Astrophys. J.} {\bf#1} (#2) #3.}
\def\apjl#1#2#3 {{ Astrophys. J. Lett.} {\bf#1} (#2) #3.}
\def\app#1#2#3  {{ Acta. Phys. Pol.} {\bf#1} (#2) #3.}
\def\ar#1#2#3   {{ Ann. Rev. Nucl. Part. Sci.} {\bf#1} (#2) #3.}
\def\cpc#1#2#3  {{ Computer Phys. Comm.} {\bf#1} (#2) #3.}
\def\err#1#2#3  {{ Erratum} {\bf#1} (#2) #3.}
\def\ib#1#2#3   {{ ibid.} {\bf#1} (#2) #3.}
\def\jmp#1#2#3  {{ J. Math. Phys.} {\bf#1} (#2) #3.}
\def\ijmp#1#2#3 {{ Int. J. Mod. Phys.} {\bf#1} (#2) #3.}
\def\jetp#1#2#3 {{ JETP Lett.} {\bf#1} (#2) #3.}
\def\jpg#1#2#3  {{ J. Phys. G.} {\bf#1} (#2) #3.}
\def\mpl#1#2#3  {{ Mod. Phys. Lett.} {\bf#1} (#2) #3.}
\def\nat#1#2#3  {{ Nature (London)} {\bf#1} (#2) #3.}
\def\nc#1#2#3   {{ Nuovo Cim.} {\bf#1} (#2) #3.}
\def\nim#1#2#3  {{ Nucl. Instr. Meth.} {\bf#1} (#2) #3.}
\def\np#1#2#3   {{ Nucl. Phys.} {\bf#1} (#2) #3.}
\def\pcps#1#2#3 {{ Proc. Cam. Phil. Soc.} {\bf#1} (#2) #3.}
\def\pl#1#2#3   {{ Phys. Lett.} {\bf#1} (#2) #3.}
\def\prep#1#2#3 {{ Phys. Rep.} {\bf#1} (#2) #3.}
\def\prev#1#2#3 {{ Phys. Rev.} {\bf#1} (#2) #3.}
\def\prl#1#2#3  {{ Phys. Rev. Lett.} {\bf#1} (#2) #3.}
\def\prs#1#2#3  {{ Proc. Roy. Soc.} {\bf#1} (#2) #3.}
\def\ptp#1#2#3  {{ Prog. Th. Phys.} {\bf#1} (#2) #3.}
\def\ps#1#2#3   {{ Physica Scripta} {\bf#1} (#2) #3.}
\def\rmp#1#2#3  {{ Rev. Mod. Phys.} {\bf#1} (#2) #3.}
\def\rpp#1#2#3  {{ Rep. Prog. Phys.} {\bf#1} (#2) #3.}
\def\sjnp#1#2#3 {{ Sov. J. Nucl. Phys.} {\bf#1} (#2) #3.}
\def\spj#1#2#3  {{ Sov. Phys. JEPT} {\bf#1} (#2) #3.}
\def\spu#1#2#3  {{ Sov. Phys.-Usp.} {\bf#1} (#2) #3.}
\def\zp#1#2#3   {{ Zeit. Phys.} {\bf#1} (#2) #3.}

\def\beq{\begin{equation}}
\def\eeq{\end{equation}}
\def\bea{\begin{eqnarray}}
\def\eea{\end{eqnarray}}

\begin{document} 
\date{\empty}


\title{\bf Suppression of Lepton Flavour Violation 
from Quantum Corrections above $M_{GUT}$}
\author{ \bf 
M.E. Gomez$^{a}$,
S. Lola$^b$, 
P. Naranjo$^a$, 
J. Rodriguez-Quintero$^{a}$
}

\maketitle

\begin{center}
$^{a)}$ Departamento de F\'{\i}sica Aplicada, University of Huelva, 
21071 Huelva, Spain \\
$^{b)}$ Department of Physics, University of Patras, 26500 Patras, Greece
\end{center}

\vspace*{2 cm}

\begin{quote}

\begin{abstract}
We study the predictions for sfermion masses and 
Lepton Flavour Violation (LFV) for the WMAP preferred parameter space 
in $b-\tau$ Yukawa-unified models with massive neutrinos. 
A soft term structure as predicted by
an Abelian flavour symmetry combined with 
$SU(5)$ RGEs for scales above $M_{GUT}$,
results to an efficient suppression of the 
off-diagonal terms in the scalar soft 
matrices, particularly for $m_0< 100$ GeV.
Using the WMAP bounds, this implies $35\le\tan\beta \le 45$, 
$350\,\mathrm{GeV}\le m_{1/2}\le 1\,\mathrm{TeV}$, with the
higher $\tan\beta$ values being favored.
Within this framework, SU(5) unification becomes
compatible with the current experimental bounds,
in contrast to the conventional case where the soft terms 
are postulated at the GUT scale. 
\end{abstract}

\vspace*{2 cm}

\begin{flushright}
{\small UHU-FP/10-023}\\
\end{flushright}


\end{quote}
\pagebreak



\section{Introduction}

The pattern of fermion masses and mixings is one of the most compelling 
mysteries in particle physics. The large hierarchies in the 
fermion mass matrices and the origin of mixing terms remain unclear.
Neutrino oscillations render this problem even more peculiar
since, data from atmospheric~\cite{skatm} and solar~\cite{sksol}
neutrinos confirms the existence of neutrino oscillations
with near-maximal  $\nu_\mu - \nu_\tau$ mixing and
large $\nu_e \to \nu_{\mu}$ mixing ~\cite{data-fits}. 

In recent years,  several attempts to explain the observed fermion 
structure have been put forward in 
the literature. Among them, flavour symmetries are particularly appealing 
\cite{FG-KR}; in these models, only third generation entries 
are  non-zero as long as the 
family symmetry remains unbroken, whereas the remaining entries are generated 
through non-renormalizable terms after symmetry breaking by fields acquiring 
non-zero vacuum expectation values (vevs).

The flavour problem is particularly challenging in supersymmetric
(SUSY) theories, where 
soft breaking terms involve off-diagonal entries and complex phases 
that may lead to unacceptably large Flavour 
Changing Neutral Currents (FCNC) and 
CP-violating vertices. Two popular solutions usually adopted in the 
literature to solve this problem are either to consider universal 
soft terms at the high scale \cite{EN} or to invoke some kind of 
alignment among the Yukawa textures and the soft terms \cite{NS}. 
Whichever  option is 
taken, however, we should keep in mind that RGE evolution from the high 
scale down to low energies also generates additional off-diagonal 
contributions,
since it is not possible 
to simultaneously diagonalise neutrino, charged lepton and 
slepton mass matrices \cite{Borzumati1}. 
These contributions also 
imply violation of the corresponding charged-lepton numbers 
~\cite{LFVhisano,LFVres, GUT, HisNo,
rpv2,reviewLFV}, generating process forbidden in the SM,
such as $\mu \to e \gamma$, $\mu-e$ conversions,
$\tau \to \mu \gamma$ and $\tau \to e
\gamma$ decays. 

As it turns out, the stringent bounds from LFV are hard to satisfy;
in fact, it was shown that models with flavour symmetries based 
on $SU(5)$ with hierarchical Yukawa textures and the lepton mixing arising 
mainly from the charged-lepton sector, tend to predict too large rates 
\cite{Leontaris,Chankowski}. 
The data from WMAP \cite{Spergel} and the resulting bounds  
on Cold Dark Matter (CDM) further 
constrain theoretical models and make 
the potential consequences of Grand Unified Theories (GUT) for Dark Matter 
worth exploring \cite{Mambrini, GLNR, EMO, Barger}.
Imposing Yukawa unification, as expected in GUTs, 
the solutions become 
even more predictive, with additional constraints on the model 
parameters \cite{ATR,GLP}. 

The purpose of this work is to study the predictions for LFV in models 
where SU(5) unification is combined with flavour symmetries, 
taking into account the RGE evolution above the GUT scale, and
focusing on the WMAP preferred area 
presented in \cite{GLNR}. We show that,
by postulating the mass matrices 
at a high scale $M_X$ and evolving them down to $M_{GUT}$, 
the pathological situation 
encountered in the conventional models \cite{Leontaris,Chankowski}
may be remedied. 
For  regions of the  parameter space with a low $m_0$, the 
pattern we end up with exhibits a sizeable suppression in 
the off-diagonal terms as compared to the textures at $M_X$, yielding 
acceptable LFV predictions. We show that this is true,
even in the case of maximal
mixing in the charged lepton sector, which is the most
dangerous one as far as LFV is concerned.
 
The paper is organized as follows: 
In Section 2 we summarise  the origin of flavour 
violation in a generic $SU(5)$ framework. In section 3 we discuss 
fermion and sfermion mass matrices in 
$SU(5)$ unification with an Abelian flavour symmetry. 
Section 4 describes the running procedure and the results. 
The conclusions are presented in Section 5.

\section{Lepton Flavour Violation in SUSY-SU(5) with see-saw neutrinos}

In SUSY theories, charged lepton flavour violation 
may be generated at the loop-level, 
even in models with universal soft 
terms at a high scale. 
This is also true for the MSSM, when
extended with a 
see-saw mechanism to generate small neutrino masses \cite{LFVhisano,HisNo}. 
In this case, LFV terms are generated 
radiatively, since it is not possible 
to simultaneously diagonalise neutrino, charged lepton and 
slepton mass matrices 
\cite{Borzumati1}. 

In order to explain the observed lepton hierarchies  by  
suitable Yukawa textures, we introduce flavour symmetries, which
may also predict the structure of the soft mass terms. 
LFV processes like 
$l_i\rightarrow l_j \gamma$ impose severe constraints on the
allowed patterns 
\cite{Leontaris,Chankowski}. 
In this work we focus on the implications of massive 
neutrinos in SU(5) Yukawa unification
with an additional $U(1)_F$ family symmetry, 
including RGE effects not only below, but also
above $M_{GUT}$.

We start by considering the following 
SUSY $SU(5)$ superpotential:  
\begin{equation}
\label{W}
{\mathcal{W}}_X=T_1^T\,{\mathcal{Y}}_u^{\delta}\,T_1\,H + 
T_1^T\,{\mathcal{\bar{Y}}}_d\,\bar{F_1}\,\bar{H} + \bar{F_1}^T\,
{\mathcal{Y}}_{\nu}^{\delta}\,S_1\,H + S_1^T\,\bar{M}_R\,S_1, 
\end{equation}
where ${\mathcal{Y}} _{\alpha}$ ($\alpha =u,d,\nu$) are the Yukawa matrices 
for the up-type quarks, down-quarks/charged-leptons and Dirac neutrinos, 
respectively. $M_R$ is the heavy Majorana mass matrix. 
The symbol $\delta$ stands 
for \emph{diagonal}, the original fields rotated as
\begin{equation}
T=U_{10}T_1,\;\;\;\;\;\;\;\bar{F}=U_{\nu L}\bar{F}_1, \;\;\;\;\;\;\;
S=U_{\nu R} S_1, 
\end{equation}
with the rotating matrices defined as
\begin{equation}
{\mathcal{Y}}_u=U_{10}{\mathcal{Y}}_u^\delta U_{10}^T,\;\;\;\;\;\;\; 
{\mathcal{Y}}_d=U_{5L}^\ast{\mathcal{Y}}_d^\delta U_{5R}^\dagger,
\;\;\;\;\;\;\; 
{\mathcal{Y}}_\nu=U_{\nu L}^\ast{\mathcal{Y}}_\nu^\delta U_{\nu R}^\dagger,
\end{equation}
and 
\begin{equation}
{\mathcal{\bar{Y}}}_d= V_{CKM}^\ast{\mathcal{Y}}_d^\delta V_E^\dagger,
\end{equation}
Here, $V_{CKM}=U_{10}^\dagger U_{5L}$ and $V_{E} = U_{\nu L}^\dagger U_{5R}$  
denote the mixings in the quark and lepton sectors, while 
$\bar{M}_R=U_{\nu R}^T M_R U_{\nu R}$.

The off-diagonal contributions to slepton mass matrices, when the 
superfields are rotated so that charged leptons become diagonal, can be 
understood through three rotations at different energy scales:
\begin{itemize}
\item $M_X$. The rotations in the superpotential fields lead 
to the following transformation of the soft terms:
\begin{equation}
\bar{m}_{10}^2=U_{10}^\dagger m_{10}^2U_{10}\;\;\;\;\;\;\;  
\bar{m}_{5}^2=U_{\nu L}^\dagger m_{5}^2U_{\nu L}
\label{rotm105}
\end{equation}
\item $M_{GUT}$. Assuming that $SU(5)$ is broken down to the 
MSSM gauge group, the superpotential becomes
\begin{equation}
{\mathcal{W}}_{MSSM}=Q^T {\mathcal{Y}}_u^\delta U H_2+ 
Q^T (V_{CKM}^\ast {\mathcal{Y}}_d^\delta) D H_2+
   L^T (V_E^\ast {\mathcal{Y}}_d^\delta ) E H_2+ 
L^T {\mathcal{Y}}_\nu^\delta S H_2+ S^T \bar{M}_R S
\end{equation}
where we have absorbed the matrices $V_E^\ast$ and $V_{CKM}^\ast$ in the 
definitions of the superfields $E$ and $D$ respectively.
The scalar soft masses then become:
\begin{equation}
m_E^2=V_{CKM}^\dagger \bar{m}_{10}^2 V_{CKM},\;\;\;\;\;
m_L^2=\bar{m}_{5}^2, \nonumber 
\end{equation}
\begin{equation}
m_Q^2=m_U^2=\bar{m}_{10}^2,\;\;\;\;\;m_D^2=V_E^\dagger \bar{m}_{5}^2 V_E, 
\end{equation}

We write $m_D^2$ and $m_E^2$ in the following way:
\begin{eqnarray}
m_D^2 & = & V_E^{\dagger}\bar{m}_5^2V_E \nonumber \\
      & = & U_{5R}^{\dagger}U_{\nu L}U_{\nu L}^{\dagger}m_5^2
U_{\nu L}U_{\nu L}^{\dagger}U_{5R} \nonumber \\
      & \simeq & U_{5R}^{\dagger}m_5^2U_{5R} \nonumber \\
m_E^2 & = & V_{CKM}^{\dagger}\bar{m}_{10}^2V_{CKM} \nonumber \\
      & = & U_{5L}^{\dagger}U_{10}U_{10}^{\dagger}m_{10}^2
U_{10}U_{10}^{\dagger}U_{5L} \nonumber \\
      & \simeq & U_{5L}^{\dagger}m_{10}^2U_{5L}
\label{options}
\end{eqnarray}
where the last step holds if radiative corrections to the rotation matrices 
are neglected.

\item $M_N$. $M_N$ is the scale at which  
the  heavy right handed neutrinos decouple. Below this scale, 
the particle content is 
just the one of the MSSM complemented with
the neutrino mass operator resulting from the see-saw mechanism. 
Consequently, the superpotential can be written in a basis where the
charged-lepton mass matrix becomes diagonal and the left slepton mass 
matrix becomes
\begin{equation}
\bar{m}_L^2=V_E^\dagger m_L^2 V_E \simeq U_{5R}^{\dagger}m_5^2U_{5R}
\end{equation}
\end{itemize}

RGE effects play a significant role in the calculation
of flavour-violating processes. Even in case of
universal soft terms at $M_X$, RGE runs between 
$M_X$ and $M_{GUT}$ (arising mainly
through superpotential terms 
of the form $\bar{E}\bar{U}\bar{H}$ where
$\bar{H}$ is a colour-triplet Higgs field)
give rise to one-loop diagrams that also renormalise
the right-handed slepton masses (which in 
the CMSSM would remain to a large extent diagonal).
In the leading-logarithmic approximation these 
corrections are given by \cite{Hisano}
\bea
(m^2_L)_{ij} 
\simeq  - \frac{3}{8\pi^2}
  \mathcal{Y}_{u_3}^2 V_{CKM}^{3i} V^{\ast 3j}_{CKM} 
 (3 m_0^2 +a_0^2)  \log  \frac{M_{\rm X}}{M_{\rm GUT}} 
\eea
for $i \ne j$, 
and, as we mentioned, are suppressed due to the smallness of $V_{CKM}$;
this holds in the minimal supersymmetric $SU(5)$, since
in extensions of the theory
this mixing may be further amplified \cite{JE-rec}. 
On the contrary, runs from 
$M_{GUT} \rightarrow M_{N}$ are crucial.
In the leading-logarithmic approximation, the non-universal renormalization 
of the soft supersymmetry-breaking scalar masses is given by 
\begin{eqnarray}
(m^2_L)_{ij} 
&\simeq& - \frac{1}{8\pi^2} \left( 
  \mathcal{Y}_{\nu_3}^2 V^{\ast 3i}_{E} V_{E}^{3j} 
 \log \frac{M_{\rm X}}{M_{\nu_3}}
+ \mathcal{Y}_{\nu_2}^2 V^{\ast 2i}_{E} V_{E}^{2j} 
 \log \frac{M_{\rm X}}{M_{\nu_2}}
\right)  (3 m_0^2 +a^2_0)  
\label{offdiagofL} 
\end{eqnarray}
implying that the corresponding corrections to left-handed slepton masses
are proportional
to $V_{E}$ (the Dirac neutrino mixing matrix in the basis where
the $d$-quark and charged-lepton masses are diagonal).
In this approach, non-universality in the soft 
supersymmetry-breaking left-slepton masses is much larger than the one in 
the right-slepton masses.

\section{$SU(5)$ textures}
Having defined the general framework, the next step consists of 
summarising $SU(5)$ Yukawa textures that match the fermion data and may
also predict the pattern of soft terms to be expected.
The mass matrices are constructed by looking at the field content
of SU(5) representations, namely: three families of
$(Q,u^{c},e^{c})_{i} \in {\tt 10}$, three families of
$(L,d^{c})_i  \in { \tt \overline{5}}$ representations, 
and heavy right-handed neutrinos in singlet representations. 
This model has therefore the following properties:
(i) the  up-quark mass matrix is symmetric, and
(ii) the charged-lepton mass matrix is the transpose of the 
down-quark mass matrix, which relates the mixing 
of the left-handed leptons to that of the
right-handed down-type quarks. 
Since the CKM mixing in the quark sector is due 
to a mismatch between the mixing of the left-handed up- and 
down-type quarks, it is independent of mixing in the lepton sector,
easily reconciling the large atmospheric neutrino 
mixing angle with the observed small $V_{CKM}$ mixing.
Following the $U(1)_F$ charge assignment in \cite{EGL},  the 
Yukawa matrices have the form
\begin{equation}
\mathcal{Y}_{u}\propto \left(
\begin{array}{ccc}
\varepsilon ^6 & \varepsilon ^5 & \varepsilon ^3 \\
\varepsilon ^5 & \varepsilon ^4 & \varepsilon ^2 \\
\varepsilon ^3 & \varepsilon ^2 & 1 \\
\end{array}
\right),\,\,\,
\mathcal{Y}_{\ell}^T\propto 
\mathcal{Y}_{d}\propto 
\left(
\begin{array}{ccc}
\varepsilon ^4 & \varepsilon ^3 & \varepsilon ^3 \\
\varepsilon ^3 & \varepsilon ^2 & \varepsilon ^2 \\
\varepsilon  & 1 & 1 \\
\end{array}
\right),\,\,\,
\mathcal{Y}_{\nu}\propto \left(
\begin{array}{ccc}
\varepsilon ^{|1\pm n_1|} & \varepsilon ^{|1 \pm n_2|} 
& \varepsilon ^{|1 \pm n_3|} \\
\varepsilon ^{|n_1|} & \varepsilon ^{|n_2|} & \varepsilon ^{|n_3|} \\
\varepsilon ^{|n_1|} & \varepsilon ^{|n_2|} & \varepsilon ^{|n_3|} \\
\end{array}
\right)
\label{Yukawas}
\end{equation}
where $n_i$ stand for the heavy Majorana neutrino charges.
As discussed in the Introduction, we will assume that
the entire lepton mixing is arising from the charged-lepton sector,
which is potentially the most
dangerous case as far as LFV is concerned. 

The rotation matrices that diagonalise $\mathcal{Y}_u$ and $
\mathcal{Y}_\ell^T \propto \mathcal{Y}_d$ 
are 
\begin{equation}
U_{10}=\left(\begin{array}{ccc}
-1+\frac{\varepsilon ^2}{2} & \varepsilon & 0 \\
-\varepsilon & -1+\frac{\varepsilon ^2}{2} & \varepsilon ^2 \\
\varepsilon ^3 & \varepsilon ^2 & 1 \\
\end{array}\right)
\end{equation}

\begin{equation}
U_{5L}=\left(\begin{array}{ccc}
-1+\frac{\varepsilon ^2}{2} & \varepsilon & 0 \\
-\varepsilon & -1+\frac{\varepsilon ^2}{2} & \varepsilon ^2 \\
\varepsilon ^3 & \varepsilon ^2 & 1 \\
\end{array}\right),\,\,\,\,\,
U_{5R}=\left(\begin{array}{ccc}
\frac{1}{\sqrt{2}} & -\frac{1}{2}-\frac{\varepsilon}{2\sqrt{2}} & 
\frac{1}{2}-\frac{\varepsilon}{2\sqrt{2}} \\
-\frac{1}{\sqrt{2}} & -\frac{1}{2}+\frac{\varepsilon}{2\sqrt{2}} & 
\frac{1}{2}+\frac{\varepsilon}{2\sqrt{2}} \\
\frac{\varepsilon}{\sqrt{2}} & \frac{1}{\sqrt{2}} & \frac{1}{\sqrt{2}} \\
\end{array}\right)
\end{equation}

While  there is no unique choice 
of the right handed neutrino charges
${n_1,n_2,n_3}$ (several choices may lead to correct 
low energy neutrino data) representative choices can
be made, and among the simplest patterns is the one provided
by the assignment $\{n_1,n_2,n_3\}=\{1,1,1\}$.
In this case, 
\begin{equation}
V_E= U_{\nu L}^{\dagger}U_{5R} =
\left(\begin{array}{ccc}
-\frac{1}{\sqrt{2}} & -\frac{1}{2}+\frac{\varepsilon}{2\sqrt{2}} & 
\frac{1}{2}+\frac{\varepsilon}{2\sqrt{2}} \\
-\frac{1}{2}+\frac{\varepsilon}{2} & \frac{1}{2}\left(1+\frac{\sqrt{2}}{2}
\right)+\frac{\varepsilon}{4} & \frac{1}{2}\left(1-\frac{\sqrt{2}}{2}\right)
+\frac{\varepsilon}{4} \\
\frac{1}{2}+\frac{\varepsilon}{2} &  \frac{1}{2}\left(1-\frac{\sqrt{2}}{2}
\right)-\frac{\varepsilon}{4} & \frac{1}{2}\left(1+\frac{\sqrt{2}}{2}\right)
-\frac{\varepsilon}{4} \\
\end{array}\right)
\end{equation}
where $U_{\nu L}$ is the left rotation matrix for the 
Dirac neutrino sector.

In addition, flavour symmetries generally imply 
non-universal soft terms \cite{reviewLFV}, since 
the structure of the soft terms
is linked to the family charges.
For the  Yukawa textures in
Eq.(\ref{Yukawas})
the soft mass matrices $m_{10}^2$ and $m_5^2$   become
\begin{equation}
m_{10}^2\propto \left(
\begin{array}{ccc}
1 & \varepsilon  & \varepsilon ^3 \\
\varepsilon  & 1 & \varepsilon ^2 \\
\varepsilon ^3 & \varepsilon ^2 & 1 \\
\end{array}
\right)m_0^2,\,\,\,\,\,
m_5^2\propto \left(
\begin{array}{ccc}
1 & \varepsilon  & \varepsilon  \\
\varepsilon  & 1 & 1 \\
\varepsilon  & 1 & 1 \\
\end{array}
\right)m_0^2,
\label{m105}
\end{equation}
The diagonalizations performed on the superfields also influence 
the soft mass terms, thus we must rotate the textures accordingly: 
\begin{equation}
\bar{m}_{10}^2=\left(\begin{array}{ccc}
1 & \varepsilon & \varepsilon ^3 \\
\varepsilon & 1 & \varepsilon ^2 \\
\varepsilon ^3 & \varepsilon ^2 & 1 \\
\end{array}\right)m_0^2,\,\,\,\,\,
\bar{m}_5^2=\left(\begin{array}{ccc}
0 & 0 & 0 \\
0 & 1 & \varepsilon \\
0 & \varepsilon & 1 \\
\end{array}\right)m_0^2
\label{m105T2}
\end{equation}

In what follows, we will analyze the predictions of the above textures
for on LFV processes 
of the type  $l_i\rightarrow l_j+\gamma$, considering that all
textures initially arise at a scale $M_X>M_{GUT}$. \\

\section{RGE runs and results}

Let us briefly discuss the running procedure. 
We use a top-down approach, the scale $M_X$ being the starting point. 
At this scale, both the Yukawa textures and the soft mass matrices are 
determined by the family symmetry charges. We evolve the \emph{3$^{rd}$ 
generation} parameters down 
to $M_{GUT}$ using the $SU(5)$ RGEs. In this 
way, both the $V_{CKM}$ and 
the $V_E$ mixing matrices are \emph{predicted}. Then, we further evolve 
the corresponding RGEs (including the right-handed neutrino 
mass scale $M_N$ and the SUSY threshold corrections) down to the electro-weak 
scale. At $M_Z$ we \emph{impose} the experimental constraints on the gauge 
couplings, as well as acceptable fermion masses and mixings. From this point, 
we employ a bottom-up approach to evolve the RGEs, using the experimental 
constraints, up to the GUT scale (properly re-obtaining the SUSY scale and 
introducing the right-handed neutrino modes). At the GUT scale we end up 
with $V_{CKM}$ and $V_E$ that are to be compared with the ones computed in the 
top-down method. Such a comparison allows us to extract information about 
the off-diagonal soft terms at the high scale $M_X$. 

As in Ref.~\cite{GLNR} the values we use are $M_X=2\cdot 10^{17}$~GeV, 
$M_N= 3\cdot 10^{14}$~GeV. The coupling $\lambda_{\nu_3}$ is determined such that  
$m_{\nu_3}\sim 0.05$~eV; $m_b(M_Z)=2.92$~GeV and $\alpha_s=0.1172$. 
The evaluation of the LFV observables is done by performing
a full diagonalization of the slepton mass matrices 
( for instance, see~\cite{LFVhisano}), inserting the full  
rotation matrices in the lepton-slepton-gaugino vertices and summing over 
all the mass eigenstates of the exchanged particles. 
Soft terms are computed in the basis where the charged leptons are diagonal. Our results 
agree with other updated estimates of the branching ratios, such as those
             given in Ref.~\cite{paradisi}.

\begin{figure}[htb!]
\begin{center}
\hspace*{-0.3 cm}
\includegraphics[width=10cm,height=8cm]{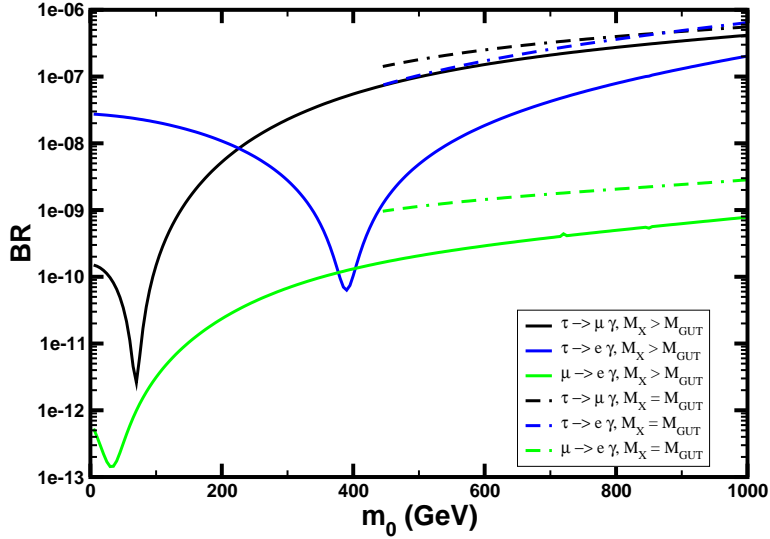}
\caption{ \it Prediction for the charged-lepton flavour violating branching 
ratios showing the difference of taking either $M_X$ or $M_{GUT}$ as the starting point
of the runs.}
\label{MXMGUT}
\end{center}
\end{figure}

\subsection{Runs above $M_{GUT}$}
The introduction of a non-trivial flavour structure for the slepton soft 
terms at $M_{GUT}$, as predicted by the 
family symmetry that also generates Yukawa couplings, 
typically results to a large violation of the bounds on $\l_j\rightarrow 
l_i \gamma$ \cite{Leontaris, Chankowski}. 
This picture may be remedied by taking into account RGE effects from a scale 
$M_X > M_{GUT}$. In this case, the cosmological requirement of 
having a neutral particle as the LSP imposes low values on $m_0$, such that 
$m_{\tilde{\tau}}>m_\chi$ \cite{Mambrini,GLNR,EMO}
(diagonal terms in the soft mass matrices have a large RGE growth, 
while  non-diagonal elements remain almost unaffected by the runs).
Thus, even assuming non-diagonal soft terms 
with matrix elememts of the same order of magnitude at $M_X$, the 
corresponding matrix at $M_{GUT}$ exhibits dominant diagonal elements. 
To some extent, the RGE effect is similar to the action of closing an umbrella: the general non-
universal soft terms at $M_X$ resemble an open umbrella that 
approaches a diagonal matrix at the GUT scale.

In Fig.~\ref{MXMGUT}, we show the differences between the following:
i) SU(5) RGE evolution of the soft terms from a high scale $M_X$ 
down to $M_{GUT}$ and then to the MSSM with see-saw neutrinos (solid lines),
and ii) Soft SUSY breaking terms given 
at $M_{GUT}$ and then the MSSM  with see-saw neutrinos (dash-lines).
In case ii) we stop the lines at the value of $m_0$ 
below which $m_{\tilde{\tau}}$ becomes the LSP. 
In contrast, $m_0$ can even vanish at $M_X$ in case i). 
The textures and soft terms we use are similar to Ref.~\cite{Chankowski}. 
However, unlike these authors,  we decouple the right-handed neutrinos 
below $M_{GUT}$. As a result, the predicted BR's do not vanish in the limit 
$m_0=0$. We also observe the presence of one 
peak for each decay; the origin of such peaks can be traced back to the 
cancellations coming from the RR sector, in agreement with \cite{paradisi}.
 
The advantageous feature of runs above the GUT scale relies on 
the increase of the mass of the lightest stau,  such that the condition 
$m_{\chi}<m_{\tilde{\tau}}$ is achieved even at low values of 
$m_0$.  These values are sufficiently low to predict rates for charged 
lepton violation  within the current experimental bounds and a relic 
density on the WMAP range\cite{Mambrini,GLNR}. 

We can provide an explicit example of the growth of the diagonal terms 
of the slepton mass matrix in models with interesting predictions for both 
LFV and $\Omega_\chi h^2$.  Let us consider the $0<m_0<100$~GeV region. 
In the area of the parameter space where the WMAP bounds are 
satisfied due to $\tau-\chi$ Co-annihilations, we find that 
$m_{1/2}$ is essentially a linear function of $m_0$, $m_{1/2}\sim a_1^i+a_2^im_0$, 
where $i$ runs over the multiplets. Taking into account 
that the radiative corrections to the off-diagonal entries of the soft mass 
matrices are subdominant as compared with those of the diagonal ones, these 
diagonal elements can be expressed as follows:

\begin{equation}
m_{S_i}^2\simeq C_i^2\left(m_0\right)m_0^2,
\label{umb}
\end{equation}
where we have defined
\begin{equation}
C_i^2\left(m_0\right)\equiv \frac{144}{20\pi}\alpha _5\left(\left(
\frac{a_1^i}{m_0}\right)^2+\frac{2a_1^ia_2^i}{m_0}+a_2^2\right)\ln\left(
\frac{M_X}{M_{GUT}}\right)
\label{MXruns}
\end{equation}
and $S_i$ stands for the supermultiplets $\mathbf{10}$ and $\mathbf{\bar{5}}$. 
As stated, Eq.(\ref{umb}) implies a large enhancement only for the diagonal 
entries of the soft matrices, further suppressing the off-diagonal 
elements. It turns out indeed that for values of $m_0\simeq 60-80$ GeV at 
$M_X$ such an enhancement at the GUT scale is as large as $\simeq$ 100. As a 
consequence, the soft mass matrices $\bar{m}_{10}^2$ and 
$\bar{m}_5^2$ at GUT scale read as
\begin{equation}
\bar{m}_{10}^2=\left(\begin{array}{ccc}
1 & \varepsilon ^3 & \varepsilon ^5 \\
\varepsilon ^3 & 1 & \varepsilon ^4 \\
\varepsilon ^5 & \varepsilon ^4 & 1 \\
\end{array}\right)C^2\left(m_0\right)m_0^2,\,\,\,\,\,
\bar{m}_5^2=\left(\begin{array}{ccc}
0 & 0 & 0 \\
0 & 1 & \varepsilon ^3 \\
0 & \varepsilon ^3 & 1 \\
\end{array}\right)C^2\left(m_0\right)m_0^2
\end{equation}
clearly exhibiting the suppression on the off-diagonal terms (as compared 
with textures (\ref{m105T2})).

\begin{figure}[htb!]
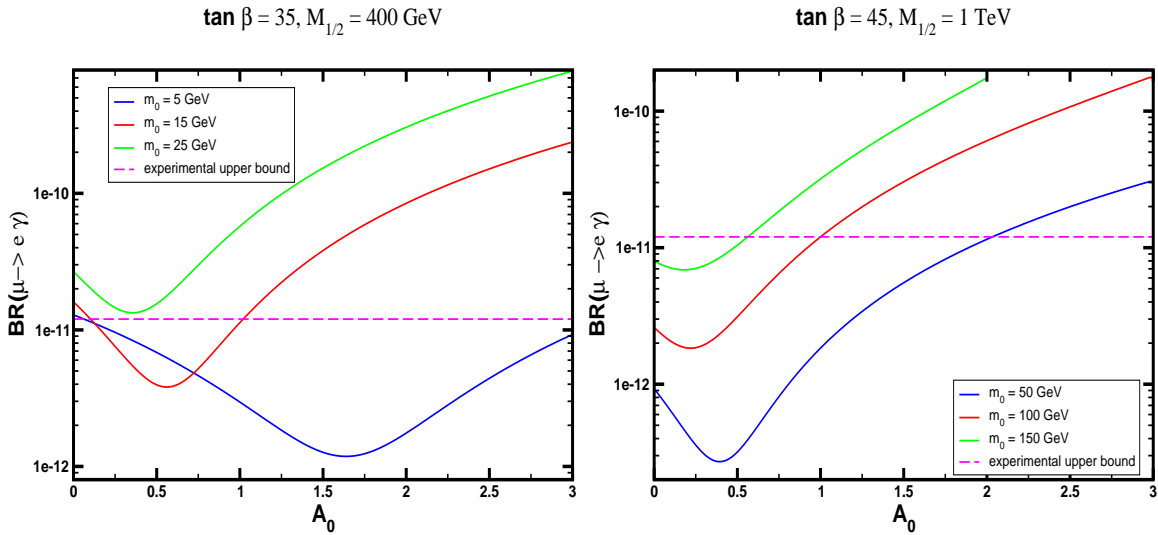

\begin{center}
\hspace*{-0.3 cm}
\includegraphics[width=7.5cm,height=7cm]{figs/BR_A0_35.eps}
\includegraphics[width=7.5cm,height=7cm]{figs/BR_A0_45.eps}
\caption{ \it Variation of the charged-lepton flavour violating branching 
ratio for $\mu\rightarrow e\gamma$ with $A_0$. 
}
\label{A0}
\end{center}
\end{figure}

Before going through the main results, it will be instructive to analyse 
the dependence of the LFV rates on the universal soft parameter $A_0$. In 
Fig.~\ref{A0}, we used again the textures and soft terms of 
Ref.\cite{Chankowski} with the first set of
right-handed neutrino charges
discussed above. We observe that, for fixed $m_0$ and $M_{1/2}$, the LFV rates 
yield 
unacceptable predictions beyond some $A_0$, hence further constraining the 
allowed parameter space (note that the allowed range for $m_0$ increases with 
$\tan\beta$). Recall that the upper limit on $A_0$ arises from the appearance 
of tachyonic soft masses.

\subsection{Runs below $M_{GUT}$ and LFV rates}

\begin{figure}[hbt!]
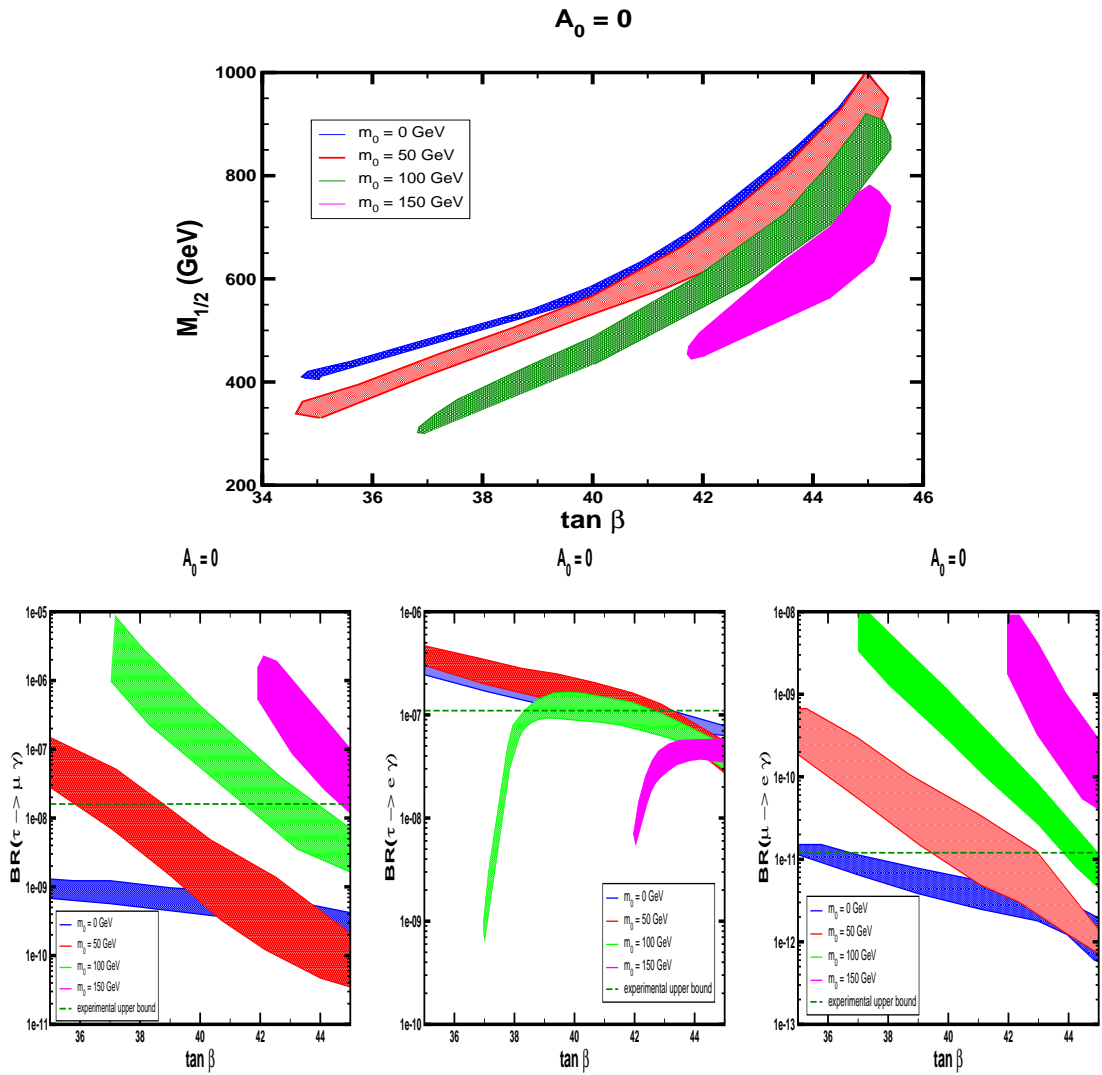

\begin{center}
\hspace*{-0.3 cm}
\begin{tabular}{c}
\includegraphics[width=10cm,height=7cm]{figs/M12tb.eps}
\\
\begin{tabular}{ccc}
\includegraphics[width=4.5cm,height=7cm]{figs/BR_tmg_tb_final.eps} &
\includegraphics[width=4.5cm,height=7cm]{figs/BR_teg_tb_final.eps} &
\includegraphics[width=4.5cm,height=7cm]{figs/BR_meg_tb_final.eps}
\end{tabular}
\end{tabular}
\caption{ \it We plot the allowed parameter space resulting from 
applying the constraints, as explained in the text, for $m_0=0,50,100,150$ 
(upper plot). The BR's in terms of $\tan\beta$ computed all along 
the allowed parameter sapce is also plotted (lower three plots).}
\label{fig:M12tb}
\end{center}
\end{figure}

An immediate question is how sensitive the results are upon variations of $m_0$.
As shown in Fig.~\ref{fig:M12tb}, the applied 
constraints imply that solutions only exist between
$\tan\beta \simeq 35$ and $\tan\beta \simeq 45$. 
Naturally, smaller values of $m_0$ lead to an enhancement 
of the allowed parameter space (while eq. 
\ref{MXruns} indicates how a higher $M_X$
enhances the allowed values of $m_0$).
Furthermore, small values of $m_0$ become favoured 
once LFV processes rates are taking under consideration  and the
``umbrella effect'' emerges. In particular, 
it can be seen that no reliable parameter combination survives above 
$m_0 \simeq 150$ GeV.

\begin{figure}[htb!]
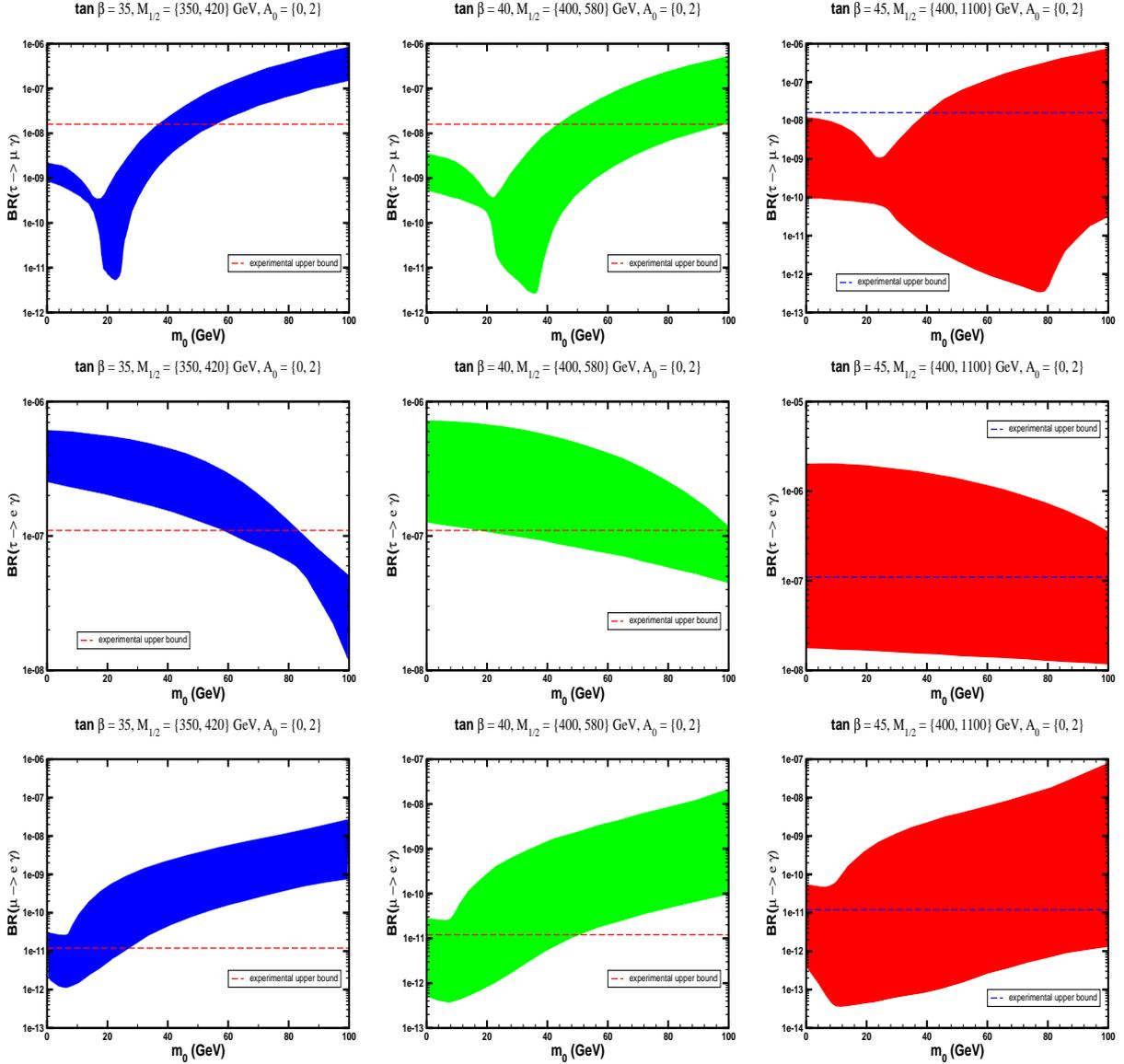

\begin{center}
\hspace*{-0.3 cm}
\begin{tabular}{ccc}
\includegraphics[width=5cm,height=5cm]{figs/BR_tmg_35_1_final.eps} 
&
\includegraphics[width=5cm,height=5cm]{figs/BR_tmg_40_1_final.eps}
&
\includegraphics[width=5cm,height=5cm]{figs/BR_tmg_45_1_final.eps}
\\
\includegraphics[width=5cm,height=5cm]{figs/BR_teg_35_1_final.eps} 
&
\includegraphics[width=5cm,height=5cm]{figs/BR_teg_40_1_final.eps}
&
\includegraphics[width=5cm,height=5cm]{figs/BR_teg_45_1_final.eps}
\\
\includegraphics[width=5cm,height=5cm]{figs/BR_meg_35_1_final.eps} 
&
\includegraphics[width=5cm,height=5cm]{figs/BR_meg_40_1_final.eps}
&
\includegraphics[width=5cm,height=5cm]{figs/BR_meg_45_1_final.eps}
\end{tabular}
\caption{ \it Prediction for the $\tau\rightarrow \mu\gamma$ (top), 
$\tau\rightarrow e\gamma$ (middle) and $\mu \rightarrow e\gamma$ 
branching ratios for the cosmologically preferred area of values 
of $M_{1/2},\,A_0$, for three different values of $\tan\beta$.}
\label{BRI}
\end{center}
\end{figure}

In the following, we shall show how the textures defined in the previous 
section yield acceptable charged-LFV rates once the umbrella effect is 
considered. As observed from Figs.~\ref{BRI} the predictions for 
the branching ratios for $\ell_i\rightarrow\ell_j+\gamma$ decays lie within 
the current experimental bounds for properly chosen parameters. 

\begin{figure}[htb!]
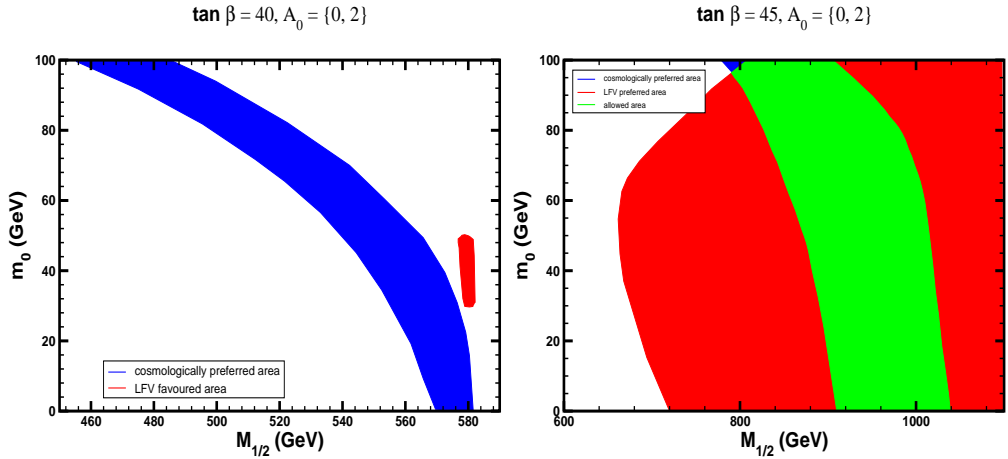

\begin{center}
\hspace*{-0.3 cm}
\includegraphics[width=6.5cm,height=6cm]{figs/m0m12_LFV_40.eps}
\includegraphics[width=6.5cm,height=6cm]{figs/m0m12_LFV_45.eps}
\caption{ \it Allowed parameter space when both LFV and cosmological 
constraints are taken into account for $\tan\beta=40$ and $45$.
}
\label{BRF}
\end{center}
\end{figure}

We should stress that the depicted ranges for 
both $m_0$ and $M_{1/2}$ are the cosmologically preferred parameter space, as 
found in \cite{GLNR}. We can see that the case $\tan\beta$ = 35 is ruled out, 
as there is no overlapping region for the three decays. This is because of the 
RR sector-induced 
cancellations mentioned above, which arises for much larger values of $m_0$ 
for $\tau\rightarrow e\gamma$ than for the other two processes. The case 
$\tan\beta$ = 40 does possess a common area for $20 < m_0 < 50$ GeV. However, 
such an area lies outside the cosmologically preferred region, as shown on 
the left panel of 
Fig.~\ref{BRF}. Thus, we conclude that the case $\tan\beta=40$ is only 
marginally allowed. Finally, for $\tan\beta$ = 45 the whole range for 
$m_0 < 100$ GeV is allowed, when suitable values for $M_{1/2},\,A_0$ are 
chosen. 
Moreover, as shown on the right panel of Fig.~\ref{BRF}, there exists an 
overlapping 
region when 
considering the cosmologically favoured parameter space 
(values of any parameters 
involved in these plots ($m_0,\,M_{1/2},\,A_0$) beyond the ranges shown lead 
to tachyonic soft masses (see also comment before Fig.~\ref{A0})). Thus, GUT 
runs efficiently suppress the off-diagonal entries, yielding charged-LFV 
rates that render the $SU(5)$ model \emph{compatible} with current 
experimental bounds.

\section{Conclusions}

In this work, we have shown that 
$SU(5)$ runs above the GUT scale, 
naturally suppress the off-diagonal entries of the soft matrices, 
through what has been called the \emph{umbrella effect}, leading to a 
nearly flavour-independent contribution for $m_0<100$ GeV. Such 
low values of $m_0$ are discarded if $SU(5)$ runs are not taken into account, 
since in this case the lightest stau becomes the LSP. Within this framework, 
$SU(5)$ runs lead to acceptable LFV rates for the cosmologically 
preferred parameter space found 
in \cite{GLNR}, favoring values of 
$\tan\beta$ around $45$. Significant deviations from this 
value, in either way, are harder to reconcile with cosmological data. 

In order to illustrate the above,
we have studied LFV predictions in the case that conventional $SU(5)$ is 
enhanced by an Abelian family symmetry.
This is in fact one of the potentially most dangerous scenarios
as far as charged-lepton flavour violation is concerned, particularly
in the simple realisations where lepton mixing (at the GUT multiplet
basis) is dominated by the charged lepton sector. 
Our results indicate that even in this case, the umbrella effect leads
to suppressions to LFV rates, leading to a very significant
enhancement of the available parameter space, as compared to
the conventional schemes, with runs below $M_{GUT}$.

Let us say a few words on future perspectives. Our next step is a more
elaborate analysis of the umbrella effect, also including flavour violating
processes from squark mixing. It would also be desirable to make a 
comparative analysis with other GUT theories, 
including \emph{left-right symmetric} models, 
which can potentially lead to further 
enhancements of the allowed parameter space.
A final point to address would be  model-dependent
features that could depend on the details of the 
heavy Majorana neutrino sector.
In this respect, flavour violating decays may shed some light to the
mass patterns of right handed neutrinos, 
which are not easily constrained by the fermion 
data alone.

\vskip 1. cm
~\\
{\bf Acknowledgements} 
We thank  John Ellis and Mirco Cannoni for useful discussions. The research 
of S. Lola and P. Naranjo has been funded by the FP6 Marie Curie Excellence
Grant MEXT-CT-2004-014297. M.E. G\'omez and J. Rodr\'{\i}guez-Quintero acknowledge support from the project P07FQM02962 funded by 
"Junta de Andalucia",
the Spanish MICINN projects FPA2009-10773 and MULTIDARK Consolider-Ingenio: CSD2009-00064.

\vskip 1. cm

{\Large{\bf Appendix }}

In this appendix we summarise the RGEs that are most relevant for the 
purposes of the work addressed in this paper. For runs above the GUT scale 
the equations involving the Yukawa couplings and the soft mass terms 
corresponding to the $\mathbf{10}$ and $\mathbf{\bar{5}}$ representations 
of $SU(5)$, for the 3$^{rd}$ generation, take the form \cite{Hisano} 

\begin{equation}
16\pi ^2\,\frac{d\lambda _N}{dt}  = \left[-\frac{48}{5}g_5^2 + 7\lambda _N^2 
+ 3\lambda _t^2 + 4\lambda _b^2\right]\lambda _N\,\, ,
\label{A11}
\end{equation}

\begin{equation}
16\pi ^2\,\frac{d\lambda _d}{dt} =  \left[-\frac{84}{5}g_5^2 + 10\lambda _d^2 
+ 3\lambda _t^2 + \lambda _N^2\right]\lambda _d\,\, ,
\label{A12}
\end{equation}

\begin{equation}
16\pi ^2\,\frac{d\lambda _t}{dt}=\left[-\frac{96}{5}g_5^2 + 9\lambda _t^2 
+ 4\lambda _d^2 + \lambda _N^2\right]\lambda _t\,\, ,
\label{A13}
\end{equation}

\begin{eqnarray}
16\pi ^2\,\frac{dm_{\mathbf{10}}^2}{dt} & = & -\frac{144}{5}g_5^2\,M_5^2 
+ \left(12\lambda _t^2 + 4\lambda _d^2\right)m_{\mathbf{10}}^2 \\ \nonumber
 &  & + 4 \left[\left(m_{\mathbf{5}}^2 + m_{\bar{h}}^2\right)\lambda _d^2 
+ A_d^2\right] + 6\left(\lambda _t^2\,m_h^2 + A_t^2\right)\,\, ,
\label{A14}
\end{eqnarray}

\begin{eqnarray}
16\pi ^2\,\frac{dm_{\mathbf{5}}^2}{dt} & = & -\frac{96}{5}g_5^2\,M_5^2 
+ 2\left(4\lambda _d^2 + \lambda _N^2\right)m_{\mathbf{5}}^2 \\ \nonumber
 &  & + 8 \left[\left(m_{\mathbf{10}}^2 + m_{\bar{h}}^2\right)\lambda _d^2 
+ A_d^2\right] + 2\left(\lambda _N^2\,m_h^2 + \lambda _N^2\,m_{\mathbf{1}}^2 
+ A_N^2\right)
\label{A15}
\end{eqnarray}

For runs from $M_{GUT}$ to $M_N$, the equations for the Yukawa matrices are 
\cite{CI}: 

\begin{equation}
16\pi^2\,\frac{d\lambda _N}{dt}=-\left[\left(\frac{3}{5}g_1^2 + 
3g_2^2\right)I_3 - \left(4\lambda _N^2+3\lambda _t^2+\lambda _{\tau}^2\right)
\right]\lambda _N\,\, ,
\label{A1}
\end{equation}

\begin{equation}
16\pi^2\,\frac{d\lambda _{\tau}}{dt}=-\left[\left(\frac{9}{5}g_1^2 + 
3g_2^2\right)I_3 - \left(4\lambda _{\tau}^2+3\lambda _b^2\lambda _N^2\right)
\right]\lambda _{\tau}\,\, ,
\label{A2}
\end{equation}

\begin{equation}
16\pi^2\,\frac{d\lambda _t}{dt}=-\left[\left(\frac{13}{5}g_1^2 + 
3g_2^2+\frac{16}{3}g_3^2\right)I_3 - \left(6\lambda _t^2+\lambda _b^2\right) 
+\lambda _N^2\right]\lambda _t
\label{A3}
\end{equation}

Since the neutrino has no coupling to the bottom quark, the Yukawa matrix 
corresponding to the latter remains unchanged with respect to the MSSM case. \\

\end{document}